\documentclass[sigconf,nonacm]{acmart}
\usepackage{graphicx}
\usepackage{epstopdf}
\usepackage{url}
\usepackage{xcolor}
\usepackage{comment}
\usepackage{subcaption}
\graphicspath{{C:/Users/dhrdesai/OneDrive - BlackRock/Papers/ICAIF_22/ArXiv_MF_Cat_NLP/images}}

\AtBeginDocument{%
\providecommand\BibTeX{{%
\normalfont B\kern-0.5em{\scshape i\kern-0.25em b}\kern-0.8em\TeX}}}

\setcopyright{acmcopyright}
\copyrightyear{2022}
\acmYear{2022}

\acmConference[ICAIF-2022]{ICAIF-2022: ACM International Conference on AI in Finance}{ 2-4 November, 2022}{New York, NY}

\begin{document}

\title{Learning Mutual Fund Categorization using Natural Language Processing}

\author{Dimitrios Vamvourellis*}
\email{dimitrios.vamvourellis@blackrock.com}
\affiliation{%
\country{BlackRock, Inc.,New York, NY, USA}
}

\author{Mate Attila Toth*}
\email{mate.toth@blackrock.com}
\affiliation{%
\country{BlackRock, Inc., Budapest, Hungary}
}

\author{Dhruv Desai}
\email{dhruv.desai1@blackrock.com}
\affiliation{%
\country{BlackRock, Inc., New York, NY, USA}
}

\author{Dhagash Mehta}
\email{dhagash.mehta@blackrock.com}
\affiliation{%
\country{BlackRock, Inc., New York, NY, USA}
}

\author{Stefano Pasquali}
\email{stefano.pasquali@blackrock.com}
\affiliation{%
\country{BlackRock, Inc., New York, NY, USA}
}

\def\Ginclude@graphics#1{%
\parpic(\Gin@@ewidth,\Gin@@eheight)[d]{#1}\picskip{0}}%
\makeatother

\renewcommand{\shortauthors}{Anonymous et al.}

\begin{abstract}
Categorization of mutual funds or Exchange-Traded-funds (ETFs) have long served the financial analysts to perform peer analysis for various purposes starting from competitor analysis, to quantifying portfolio diversification. The categorization methodology usually relies on fund composition data in the structured format extracted from the Form N-1A. Here, we initiate a study to learn the categorization system directly from the unstructured data as depicted in the forms using natural language processing (NLP). Positing as a multi-class classification problem with the input data being only the investment strategy description as reported in the form and the target variable being the Lipper Global categories, and using various NLP models, we show that the categorization system can indeed be learned with high accuracy. We discuss implications and applications of our findings as well as limitations of existing pre-trained architectures in applying them to learn fund categorization.
\end{abstract}



\maketitle
\section{Introduction}

Mutual funds and Exchange Traded Funds (ETFs) in a nutshell are baskets of securities (stocks and bonds) either tracking a market index and hence replicating the performance of the index, or attempt to outperform a chosen market index. They have become popular investment vehicles for both retail as well as institutional investors for the diversification of the portfolio they may provide over individual securities.

With ever growing popularity of funds and many market indices as well as strategies to outperform them to choose from, it is no wonder there exist a plethora of funds from different fund managers in the market. Due to the wide variety of funds available to the investor to choose from, various third-party data vendors have come up with peer group categorization systems to compare and contrast different funds according to their portfolio composition. Some of the most widely used fund categorization systems include Lipper \citep{lippercategory} and Morningstar \citep{morningstarcategorization} categories. Here, each mutual fund in the catalogue is assigned a unique category and is revisited by the experts at a regular interval such as every month. 

These categorization systems go deeper than the broader asset class based classification (equity, fixed income, etc) and provide further granular categories based on the portfolio breakdown. They have been used to identify the top performing as well as worst performing funds within their peer groups, called peer analysis of funds; to identify a home-grown fund to recommend against a competitor's fund; to explain similarities and advantages of home-grown products compared to competitors' products for marketing purposes; to quantify portfolio diversification of a given fund of funds; etc.

Such categorization systems have been investigated using machine learning (ML) techniques all the way since 1990s. Most of the earlier works such as Ref.~\cite{marathe1999categorizing,kim2000mutual,castellanos2005spanish,moreno2006self,acharya2007classifying,lajbcygier2008soft,haslem2001morningstar,sakakibara2015clustering,lamponi2015data,menardi2015double,vozlyublennaia2018mutual}attempted to reproduce the then available Morningstar categories by using an unsupervised clustering such as K-means, self-organizing maps, etc. on fund composition and returns related variables. The 'miss-clustered' funds, (with respect to the categories if considered as the ground truth labels) between the categories and the clusters were argued to suggest potential inconsistencies in the categorization methodology. The reader is referred to \cite{desai2021robustness} for a recent review on the earlier attempts.

Then, in a comment by Gambera, Rekenthaler and Xia to Ref.~\cite{haslem2001morningstar}), it was argued that the variables used for clustering in Ref.~\cite{haslem2001morningstar} were not exactly the variables Morningstar used for their categorization, and that a categorization system is by definition a \textit{classification} system so a clustering method may not be able to reproduce it. Pivoting with these counterarguments, recently, the authors in Ref.~\cite{mehta2020machine} reformulated the problem as a supervised multi-class classification problem with only fund composition related variables as the input variables. Then, an ML technique could approximate the mapping between the input and target variables with very high accuracy yielding that the categorization was indeed an internally consistent and rule-based methodology.

Though this classification still did not resolve the mystery why earlier results could not reproduce the Morningstar categories using unsupervised methods if supervised methods could reproduce the system with high accuracy. The mystery was resolved in Ref.~\citep{desai2021robustness}, where the authors utilized distance metric learning to reverse engineer the underlying distance metric implicitly used by Morningstar for Global Categorization, and then used this distance metric to perform K-means clustering. Then, the categorization system was completely reproduced even by a clustering method when the new distance metric was used.

The closest work to the current one is Ref.~\cite{satone2021fund2vec} where data for U.S. domiciled equity index funds was translated as a bipartite network where funds were represented as nodes, stocks as edges and portfolio weights as edge weights. In particular, a graph machine learning algorithm, Node2Vec \cite{grover2016node2vec}, was used to learn embedded representation of the bipartite network to identify similar funds and perform further analysis.

All the analysis in the aforementioned works has been performed only on structured data. In the present work, we propose to investigate mutual fund categorization by directly using textual data as reported to Securities and Exchange Commission (SEC) with the help of natural language processing (NLP). This novel approach has multiple advantages. First, we verify if the raw unstructured data as recorded in the SEC filings indeed has enough information to categorize a fund accurately with respect to an expert-driven rule-based categorization system. With the help of local explainability methods such as Shapley values, we can identify important features to classify a fund to a specific category (as opposed to another category). Next, along the way of training a model to learn the categorization system, as a byproduct, we also learn the embedded representation of the investment strategy text in a supervised matter, which in turn may be used to compute various other quantities such as similarity among funds or may be used as an input to other models for related tasks (e.g., predicting returns of mutual funds). In addition, once trained, in theory, the machine then can also be employed to classify a completely new fund to closely mimic the expert-driven rule-based system.

Other works that come closer to our work in spirit are related to NLP analysis of various filings related to companies. In Ref.~\cite{ito2020learning}, a BERT model was fine-tuned to industry sectors to learn embedding of companies from annual reports. In \cite{lamby2018classifying}, a Word2Vec model was used to learn embeddings of companies from news articles dataset. Ref.~\cite{hirano2019related}, Word2Vec model was trained using news and Wikipedia articles and companies' official disclosure files to identify embeddings for Japanese companies to select stocks for themed funds. To the best of our knowledge, the present work is the first work analyzing mutual funds filing data using NLP techniques.

\section{Data Description}

For the scope of this work, we used data from Lipper Global Data Feed. We used all U.S. domiciled open-end mutual funds and ETFs. SEC regulations require that a fund's prospectus must contain its principal investment objective. We extracted this self-reported principal investment objectives text for all the funds. Funds may frequently be capital appreciation, income, or both. The funds seeking capital appreciation primarily invest in assets which the fund expects to increase in value. Funds seeking income primarily invest in securities that produce income, such as bonds paying interest or securities which pay dividends. The text  often mentions benchmarks or indexes the fund is trying to track, attempting to outperform or at least compare its performance against. 

For the target variable, we used Refinitiv Lipper Global categories classifications system. Refinitiv Lipper originally grouped funds together based on their prospectus-objectives. Their classification scheme is based off Refinitiv Lipper's holding’s-based classification model and granular peer groupings which work in tandem with their legacy objective \cite{lippercategory}. There was a total of 110 categories (classes) in the dataset. After removing all rows with any missing values and setting minimum samples for stratification (9 samples per class) we had a total of 94 classes. The cleaned dataset had a total of 12594 samples. On an average there were ~304 characters for a fund's investment strategy with a standard deviation of 54. In summary, in the ML language, the problem of learning fund categorization is a multi-class classification problem with the input features being the principal investment objective text and the target variable being the Lipper Global categories with highly imbalanced dataset.

\section{Methodology}
To frame fund category prediction from text as a supervised learning problem, the character sequence representing each input observation needs to be encoded as a numeric vector of fixed length. As such, one key consideration for text classification is the choice of the underlying representation used. Traditionally document representations have been count-based, encoding word frequencies within a document as a high-dimensional sparse vector with length equal to the number of words in the vocabulary. In addition to resulting in sparse high-dimensional vectors, this 'bag-of-words' approach also discards information on term sequence and document structure. Despite its shortcomings, count based representations such as the bag-of-words or TF-IDF approach are still common and have been used in many applications including financial text analysis with good results. More recently distributed text representations -- often called embeddings -- have been introduced which encode text as shorter dense vectors and tend to perform better at capturing semantics. In the past decade advances in embedding techniques -- from static embeddings such as Word2Vec \cite{mikolov2013} to contextual embeddings like BERT \cite{devlin2018bert} and variants -- have lead to multiple breakthroughs in NLP. Given the importance of the underlying representation for classification, we've explored several methods detailed in \ref{sec:models}.

\subsection{Data Preprocessing}
Prior to using the investment objective descriptions to predict the fund categories, a sequence of preprocessing steps was performed on the raw text. Our pipeline includes common text preprocessing methods, namely tokenization, stop-word filtering, lowercasing, n-gramming and lemmatization. Most often, the first step in a text preprocessing pipeline is converting the raw text string into a sequence of tokens. This 'tokenization' process is achieved by splitting the string on non-alphanumeric delimiters e.g. whitespace or punctuation. Stop-word filtering is used to remove frequent words (e.g., articles, prepositions) that are not topical and don't carry much information with respect to the classification. Given that upper and lowercase word forms most often share the same meaning uppercase forms are converted to lowercase. N-gramming is applied to convert sequences of $n$ terms into single tokens to account for context, e.g., 'machine learning'. Lemmatization is used to normalize words with the same root into a single form, for example, 'housing' and 'houses' would be mapped to the single form 'house'. 

Given that the goal of preprocessing is to prepare text data for classification, it is dependent on the downstream architecture, namely the feature extraction and classification method used. As such not all models use the same preprocessing. For TF-IDF vectorization, we used Spacy's \cite{spacy2} 'en\textunderscore core\textunderscore web\textunderscore lg' model for tokenization, removal of stopwords and non-alphabetic sequences, as well as lemmatization. We used the 'ngram\textunderscore range' parameter of the TFidfVectoriser from the \textit{Sci-kit Learn} \cite{scikit-learn} package to generate both unigrams and bigrams. Given that TF-IDF results in high dimensional sparse feature vectors, heavier preprocessing is applied to constrain the vocabulary and consequently the dimensionality of the vectors. For the Word2Vec and Doc2Vec approaches, only tokenization and lowercasing are applied, also using Spacy. 
For BERT, we use the standard BERT-base-cased tokenizer which tokenizes input text using WordPiece and a vocabulary size of 30,000 tokens. For FinBERT we used the model-specific tokenizer, which is based on FinVocab, a vocabulary generated from financial corpora using the SentencePiece library.

\subsection{Training-testing Split (Stratified)}
To test the performance of every model on out-of-sample data we set 25\% of the data aside as test set. We further split the remaining 75\% of data to a training (85\%) and a validation set (15\%). Additionally, because of the imbalanced natured of the dataset at hand, we used stratified splits to ensure almost identical distribution of the data over all the categories in the training, validation and test set as in the original dataset.

\subsection{Models}
\label{sec:models}
\subsubsection{TF-IDF}
One of the major limitations of the bag-of-words approach is that raw word frequencies are often not very discriminative. Frequent words like 'the', 'it', 'for' are not informative of what a document is about. One way to address this is to construct document vectors using TF-IDF scores instead of the raw word counts. The TF-IDF score is a statistic to measure how relevant a word is to a document in a corpus. TF-IDF score is constructed from the term-frequency (TF) and the inverse-document frequency (IDF) statistic as shown in equations below:
\begin{equation}
\nonumber
\label{eq:tfidf}
TF{\text -}IDF_{t,d}=TF_{t,d}\times IDF_{t} \textbf{;}\; 
TF_{t,d} = \frac{n_{t,d}}{n_d} \textbf{;} \;
IDF_{t} = log_{10}\left(\frac{N}{N_t}\right),
\end{equation}
where $n_{t,d}$ stands for the number of times term $t$ appears in document $d$, $n_d$ stands for the total number of terms in document d, $N_t$ refers to the number of documents that contain the term $t$ and $N$ is the total number of documents in the corpus. 

\subsubsection{Word2Vec}
\label{sec:word2vec}
Word2Vec \cite{mikolov2013} is an algorithm introduced by Mikolov et al. to create dense word embeddings. One major advantage of this method is that it can be trained in a self-supervised manner on the raw text with no human annotation required. The core intuition behind the algorithm is that words that occur in similar contexts tend to have similar meanings -- often referred to as the distributional hypothesis. Given a context window, Word2Vec tries to predict the context words from the target word (cbow) or vice versa (skip-gram) using a shallow neural network. In this work we use a 300 dimensional Word2Vec model trained on Google news dataset. 

\subsubsection{Doc2Vec}
Doc2vec \cite{quoc2014} is a self-supervised algorithm used to create embeddings for word sequences such as paragraphs or entire documents. Doc2vec extends the word2vec approach of learning distributed representations of words to documents, where embeddings for words and documents are jointly learned. Similar to Word2Vec, Doc2Vec also comes in two flavors for computing embeddings: dbow and dpmv. The dbow or distributed-bag-of-words model is analogous to the skip-gram variant of word2vec. However, instead of using the center word to predict the context words in each context window as in the original skip-gram architecture, dbow uses a document id to predict randomly sampled words from the document. In this work we use gensim implementation of the dbow variant of Doc2Vec with 40 epochs, context window of 15 tokens and vector dimension of 100.

\subsubsection{BERT-finetuned}
BERT \cite{devlin2018bert} (Bidirectional Encoder Representations from Transformers) is a language model consisting of a stack of multiple Transformer \cite{vaswani2017attention} blocks. The model is trained in a self-supervised fashion, on the task of Masked Language Modeling. On each sentence of the training set, 15\% of the tokens are randomly masked. A softmax layer over the vocabulary size which is stacked on top of the last encoder layer, is then trained together with the Transformer layers to predict the masked words. Also, the model is further pre-trained on the task of next sentence prediction, in which the model is fed with two sentences and is trained to predict whether the second sentence follows the first. 

The first token of every sequence fed to the model is always the special classification token [CLS]. The final hidden state
corresponding to this token is used as the aggregate representation of the entire sequence for classification tasks, like the one pursued here. 

In this work, we use BERT-base Cased model as the base and we further finetune it on the task of predicting Lipper Global categories, using the fund description as the input text. Specifically, we stack a softmax layer of 94 dimensions (equal to the number of distinct Lipper Global categories) on top of the pretrained BERT-base model. As it
was pointed out by Howard and Ruder \cite{howard2018universal} fine-tuning BERT on a new task can quickly cause model to "forget" the information learned from language modeling task in the process of adapting to the new task. To prevent "catastrophic forgetting", we apply gradual unfreezing during finetuning. We first freeze all layers apart from the softmax layer which we train for 15 epochs using a high learning rate of 0.01, mini-batch size of 16 and maximum sequence length of 75 tokens. We then unfreeze all layers and further train the entire model for another 5 epochs using a smaller learning rate of $2e-5$ to prevent base layers from forgetting basic language information while focusing on this classification task. The model was finetuned on a n1-highmem-32 GCP instance with 2 TESLA T4 GPUs, 32 CPUs and 208 GB of host memory. The 20-epoch finetuning routine was completed within approximately 21 minutes (wall-clock time) using the aforementioned machine configuration.

\subsubsection{SBERT} Sentence-BERT \cite{reimers2019sentence} is a modification of BERT which uses siamese and triplet network architecture to derive semantically meaningful sentence embeddings, setting a new state-of-the-art performance on semantic similarity tasks. In this work, we test the quality of pre-trained sentence embeddings on a classification task. Specifically, we use SBERT which has been finetuned using MiniLM-L12-H384-uncased \cite{wang2020minilm} model as a base and finetuned on 1 billion sentence pairs on the task of semantic similarity. Using SBERT we map each fund description to a 384-dimensional dense vector space. We then use these embeddings as features and train a logistic regression classifier to predict Lipper Global Categories. 

\subsubsection{FinBert}
FinBERT \citep{yang2020finbert} is a finance domain-specific BERT model. The authors further finetuned BERT-base model using the original Masked Language Modeling technique on a large financial corpus of 4.9 billion tokens consisting of SEC corporate filings, earnings transcripts and analyst reports. Using FinBERT as a base, we apply the same finetuning procedure as the one described in the BERT section above and further train this model on the task of Lipper Global category classification. In this experiment, we test whether a model which has been further pre-trained on a relevant language modality can achieve superior performance to the finetuned general-use BERT model on this classification task. The results for all of the above models are summarized and discussed in section \ref{Results}. We use the same training parameters and machine configuration as the ones mentioned for BERT-finetuned above. The 20-epoch finetuning routine was completed within approximately 19 minutes (wall-clock time).

\subsubsection{Logistic regression} Most of the vectorization method detailed above --namely TF-IDF, Word2Vec, Doc2Vec and SBERT-- were used as features for a multinomial logistic regression classifier fit to Lipper Global Categories as the target variable. For BERT base and FinBERT models we did not need an additional classifier as for these models pre-training/fine-tuning approach is used. For the logistic regression, 'L2' regularization was applied with the penalty parameter tuned on a validation set with grid-search. 

\subsection{Evaluation Metrics}
The target variable in the present dataset consists of multiple classes and is highly imbalanced. Hence, we used various metrics to evaluate the performance of the above models:

\subsubsection{Weighted Accuracy} 
Accuracy is defined as the fraction of correct model prediction. For multi-class classification problems such as the present one, we use the weighted accuracy as below \cite{scikit-learn}: 
\begin{equation}
\notag
accuracy(y,\hat{y}) = \sum_{j=1}^{C} w_j \sum_{i=1}^{n} 1(\hat{y}_i=y_i),
\end{equation}
where $1(x)$ is the indicator function; $C$ is the number of classes; $w_i$ is the weight assigned to the $i$-th class such that $\sum_{i=1}^{C} w_i = 1$; $\hat{y}_{i}$ is the predicted value of $i$-th sample and $y_i$ is the corresponding true values; and $n$ is the total number of data-points.

\subsubsection{Weighted F1 score}
The unweighted F1 score is defined as $\frac{2(Precision \times Recall)}{Precision +\,Recall}$, where 'Precision' is $TP/(TP + FP)$ and 'Recall' is $TP/(TP+FN)$, with TP is the number of true positives, FP is the number of false positives and FN is the number of false negatives. However, since the learning task at hand is a multi-class classification problem, we use the micro F1 score, which favors all the classes equally, and the macro F1 score, which calculates the F1 score for each label and then take their unweighted sum, as implemented in \emph{Sci-kit Learn} \cite{scikit-learn}.

\subsubsection{AUC-ROC}
The Receiver-operating-characteristic (ROC) curve is a plot which usually yields the performance of a binary classifier system as a function of the probability discrimination threshold: the curve is generated by plotting the fraction of true positive rate (TPR) vs. the fraction of false positive rate (FPR). 
Then, the area under the ROC curve (AUC-ROC) is the probability that a classifier will rank a randomly chosen positive instance higher than a randomly chosen negative instance. For a multi-class classification problem one can follow the one-vs-rest strategy, i.e., the average of the AUC-ROC for each class against all other, or the one-vs-one strategy, i.e., averaging the pairwise AUC-ROC \cite{10.1016/j.patrec.2005.10.010,10.1023/A:1010920819831}. Again, we used micro and macro versions of the AUC-ROC to take into account the imbalanced-ness of the dataset \cite{scikit-learn}.

\subsubsection{Top-k Accuracy}
The top-k accuracy score generalizes the usual accuracy score: a model predicts probabilities for a data-point to belong to each class, and usually the class with the highest probability among all is selected as the predicted class for the data-point. However, for multi-class problems, the classes with consecutive probabilities in descending order may also be worth considering. For the top-k accuracy score, a prediction is considered correct if the ground truth label is among the k highest predicted probabilities. Formally, it is defined as 
\begin{equation}
\mbox{top-k accuracy}(y,\hat{y}) = \frac{1}{n}\sum_{i=1}^{n-1}\sum_{j=1}^{k} 1(\hat{y}_{i,j}=y_i),
\end{equation}
where $\hat{y}_{i,j}$ is the predicted class for the $i$-th data-point corresponding to the $j$-th largest predicted probability score.

\subsubsection{Confusion Matrix} 
Confusion matrix is the matrix $M$ where the element $M_{ij}$ is equal to the number of data-points that belong to the $i$-th class but are predicted to be in the $j$-th class by the given model. The elements of $M$ can be normalized by the sum of each row.

\subsection{Explainability}\label{section:explainability}
Simple interpretable models are favored in many fields where model explainability is crucial for downstream decision making. In NLP, however, the superior performance of complex models with millions of parameters like BERT justifies their use in real-world applications over simpler, more interpretable models. Yet, we can still explain predictions and perform feature attribution even for black-box models like BERT. In this work, we explore the quality and usefulness of explainability methods using the predictions of the model with the highest out-of-sample performance on fund classification, namely \textit{BERT-finetuned}. Our goal is a) to better understand the most important features, in this case words, which drive BERT's decisions and b) to uncover limitations of the model and modes of systematic error. 

To do this, we employ the Partition explainer implementation of the SHAP package \cite{NIPS2017_7062}, which is based on Shapley values \cite{shapley1997value}. At a high-level, Shapley values provide a model-agnostic method to interpret machine learning model predictions. The Shapley value is equal to the contribution of a feature to the final prediction - Shapley values for all features add up to the final prediction for a given data point. In the NLP setting, Shapley value is the contribution of each token in the input sentence to the model's prediction, in this case the log-odds of each Lipper category class. We think of each sentence as the set of tokens $F$. For each token $i$, we first compute the marginal contribution of $i$ to the model output function $v$ for all subsets $S$ in $F$ that exclude $i$, which is equal to the difference between the model output with $i$ and the output without appending $i$ to $S$. Shapley value for token $i$ is the weighted marginal contribution over all subsets $S$, where greater weights are given to subsets that are close to either the initial set of tokens or the empty set. Mathematically, the Shapley value for token $i$ is defined as
\begin{equation}
\phi_i(v, F) = \sum_{S \subseteq F - \{i\}} \frac{\lvert S \rvert !(\lvert F \rvert - \lvert S \rvert - 1)!}{\lvert F \rvert !} [v(S \cup \{i\}) - v(S)]
\end{equation}
\vspace{-4mm}

Thus to compute the Shapley value for token $i$ in a sentence with $|F|$ tokens, we have to perform $2^{\lvert F \rvert}$ computations. Clearly, this is not tractable as the length of the input text increases. Instead, the Partition explainer uses Owen values \cite{owen1977values} which approximate Shapley values by reducing the number of subsets on which the marginal contributions for each token are computed. At a high level, the marginal contributions are calculated based on coalitions, which are unions of tokens that respect the order of words in the original sentence. 

\begin{figure*}[!htbp]

\begin{subfigure}{0.9\textwidth}
\includegraphics[width=0.9\textwidth]{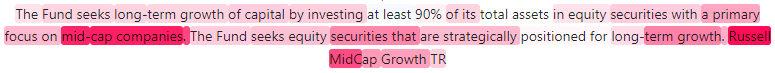}
\caption{Local feature importance for Equity US Sm \& Mid Cap}
\label{fig:local_expl_eq_midcap}
\end{subfigure}

\begin{subfigure}{0.9\textwidth}
\includegraphics[width=0.9\textwidth]{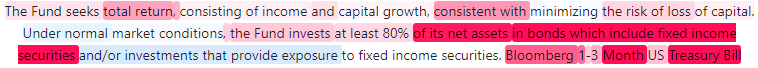}
\caption{Local feature importance for Bond USD Short Term}
\label{fig:local_expl_bond_usd_short}
\end{subfigure}
\caption{Explaining local feature importance using Owen values. Here the color scheme ranges from red to blue, where red indicates high and blue indicates low feature importance.  }
\end{figure*}

\begin{table*}[!htbp]
\centering
\begin{tabular}{*9c}
\toprule Model & \multicolumn{4}{c}{Top-k Accuracy} & \multicolumn{2}{c}{F1-Score} & \multicolumn{2}{c}{AUC} \\
\midrule
{} & k=1 & k=2 & k=3 & k=4 & Micro & Weighted & Micro & Weighted \\
\textbf{BERT-finetuned} & \textbf{0.745} & \textbf{0.856} & \textbf{0.900} & \textbf{0.923} & \textbf{0.740} & \textbf{0.744} & \textbf{0.994} & \textbf{0.986} \\
\textbf{TF-IDF+LR} & 0.670 & 0.805 & 0.851 & 0.875 & 0.670 & 0.659 & 0.984 & 0.973 \\
\textbf{Word2Vec+LR} & 0.608 & 0.758 & 0.820 & 0.851 & 0.608 & 0.601 & 0.985 & 0.969 \\
\textbf{Doc2Vec+LR} & 0.495 & 0.671 & 0.745 & 0.789 & 0.496 & 0.474 & 0.976 & 0.942 \\
\textbf{SBERT+LR} & 0.680 & 0.813 & 0.868 & 0.897 & 0.680 & 0.666 & 0.990 & 0.977 \\

\textbf{FinBERT} & 0.744 & 0.856 & 0.897 & 0.922 & 0.738 & 0.744 & 0.993 & 0.986 \\

\bottomrule
\end{tabular}
\caption{Out-of-sample performance metrics for BERT-finetuned and baseline models (Logistic Regression (LR)). BERT surpasses baselines both on aggregate accuracy as well as in terms of metrics which take class imbalance into account.}
\label{table:acc}
\vspace{-4mm}
\end{table*}

Using Owen values in this context, we can attribute the change in the log-odds of a given class to each token of the input fund description. Hence, we are able to perform local explanations by looking at the Owen value of each word in a given fund description of the test set and understand which words the model mostly attended to when making the final prediction. Summing up the Owen values from all test points for each class, we can also perform global explanations. Tokens with the highest Owen value sums are the words which increased the log-odds for this class across the whole test dataset, and in turn these words are the most important features for this class. In section \ref{Results} we show local explanations for some misclassifications as well as the most important words for some of the most frequent classes.

\begin{table*}[!htbp]
\centering
\begin{tabular}{llll}
\toprule
\textbf{EQUITY US} & \textbf{EQUITY US SM \& MIDCAP} & \textbf{EQUITY SECTOR INFO TECH} & \textbf{EQUITY EM MKTS GLOBAL} \\
\midrule
1000 & Russell & technology & emerging \\
equity & 2000 & Technology & market \\
500 & - & Global & markets \\
stocks & companies & infrastructure & erging \\
growth & small & internet & em\\
\toprule
\textbf{BOND USD MUNICIPAL} & \textbf{BOND USD MEDIUM TERM} & \textbf{BOND USD SHORT TERM} & \textbf{MONEY MARKET USD} \\
\midrule
Municipal & Bloomberg & Bloomberg & liquid \\
Federal & Barclays & Barclays & ity \\
tax & bonds & 3 & preservation \\
municipal & debt & debt & Treasury \\
preservation & gate & 1 & level\\
\toprule
\end{tabular}
\caption{BERT-finetuned feature importance - top 5 words with highest Owen value sums (i.e. measure of global feature importance) on test set for 12 sample classes.}\label{table:global_explan}
\vspace{-2.5mm}
\end{table*}

\begin{figure*}[!htbp]
\centering
\includegraphics[width=0.75\linewidth]{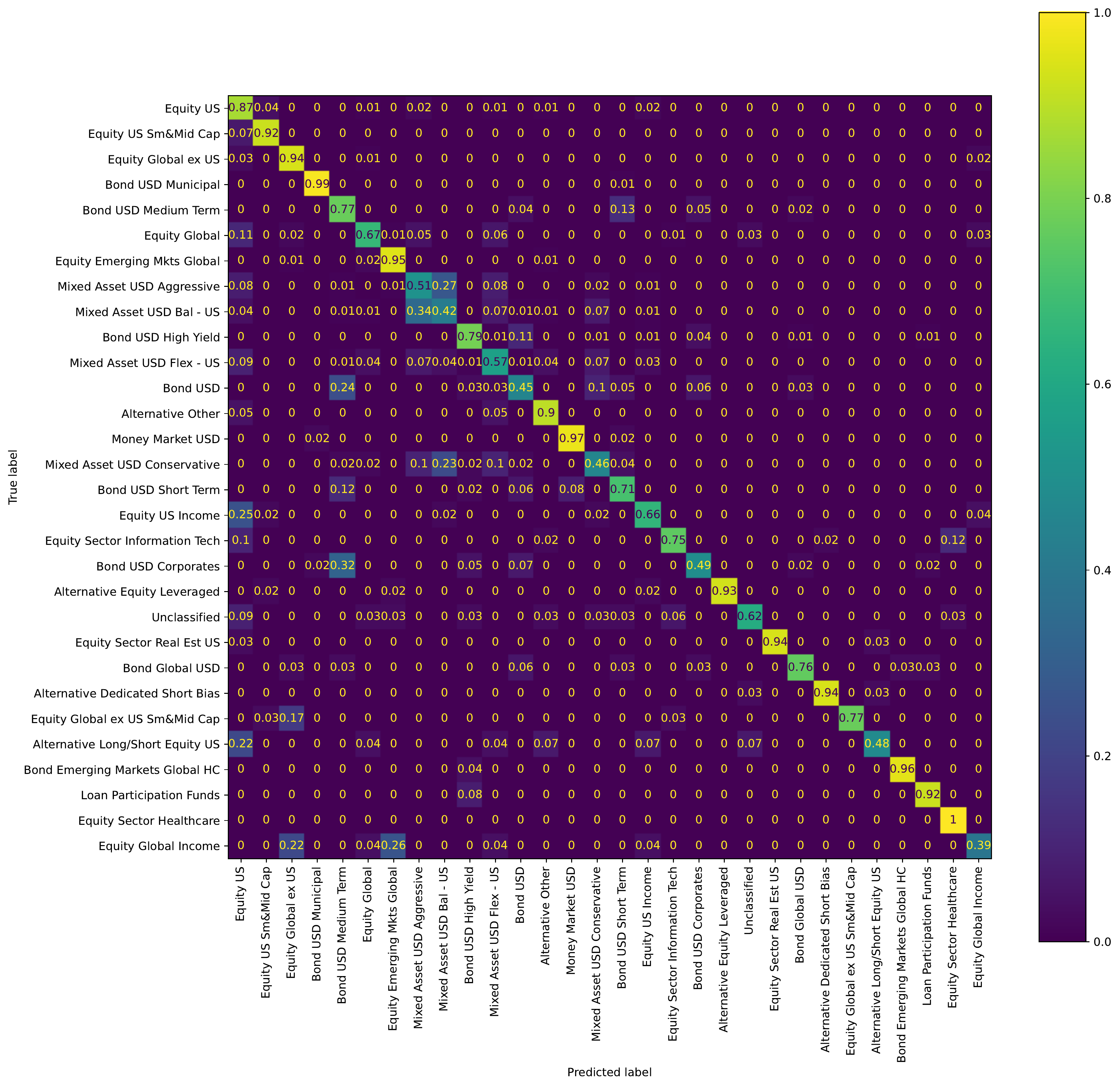}
\caption{Confusion matrix for 30 most frequent classes on test set for BERT-finetuned.}
\label{fig:conf_matrix}
\end{figure*}

\section{Results and Discussion} \label{Results}
\textbf{BERT-finetuned outperforms baselines.} Table \ref{table:acc} summarizes performance of classifiers on out-of-sample data. BERT, which is finetuned on this classification task significantly outperforms all baselines in every metric reported. The results demonstrate the effectiveness of transfer learning and the use of pre-trained language models for a downstream task, which in this case is text-based fund classification. General-purpose BERT achieves state-of-the-art performance after finetuning on a small training set of approximately 8000 data points for 20 extra epochs, using the knowledge learned from the massive corpora that the base model was pre-trained on. On the contrary, Doc2Vec embeddings were not descriptive enough for predicting fund labels. It seems that the small size of the training set does not allow for learning descriptive Doc2Vec representations from scratch. 

Document embeddings created through word vector averaging using the pre-trained Word2Vec model described in Section \ref{sec:word2vec} yielded significant improvements over Doc2Vec. This highlights that leveraging pre-trained models (either directly as features or through fine-tuning) is beneficial when applying semantic representations on small datasets. Interestingly, aside from BERT-based models, TF-IDF vectorization performed best, outperforming both Doc2Vec and pre-trained Word2Vec approaches. This is likely due to the fact that the presence of certain words or bigrams could be highly indicative of particular funds and these are captured directly as features using a local representation such as TF-IDF. Individual terms might not be captured as well using Word2Vec averaging and Doc2vec approaches, as these create distributed representations of the entire document. 
One significant advantage of BERT is that one can finetune the base language model at the same time with the softmax classification layer, thus optimizing the entire model end-to-end for the given task. In other words, by supervising BERT using the labels from the training set, we also optimize the hidden layers of the model which are used to extract features which are particularly useful for the classification task in hand. Contrary to this, for other baseline models like SBERT, TFIDF and Doc2Vec, we first use the NLP model to perform feature extraction and then we use these features to fit a Logistic Regression classifier to perform classification. As a result, the feature extraction is not supervised by the labels as in the case of BERT, which results in significantly lower performance. 

Finally, as it is indicated by the micro F1-score which takes class imbalance into account, BERT maintains the highest performance both for the high-frequency and low-frequency classes, which is an important consideration given how imbalanced the classes are in this classification task. Also, BERT predicts the fund label with approximately 75\% accuracy, however, in about 86\% of the cases the true label is one of two classes with the highest predicted probability, as indicated by top-2 accuracy metric.

\textbf{Using a domain-specific BERT base does not further improve performance.}
We investigated the effect of applying the same finetuning process used for BERT on a different base model, FinBERT, which was further pre-trained on a financial domain-specific corpus. Having been finetuned on SEC corporate filings, the hypothesis was that FinBERT might capture the esoteric language used in financial documents and better represent words which might have a different meaning in financial jargon (e.g. bonds, debt, alternatives). As shown in table \ref{table:acc}, FinBERT achieves almost identical performance to the general-purpose BERT when finetuned on this classification task. Our theory, is that BERT can easily adapt after a couple of training epochs to the language of this financial corpus, learning the features which are particularly useful for this classification task. Thus, there is not much room for improvement that further pre-training on a relevant corpus could achieve.

\textbf{BERT feature importance.}
Using the framework described in \ref{section:explainability}, we calculated the Owen value of each token in every input fund description of the test set and for each unique class, a measure of the local importance of a token in classifying a specific text piece. In Figures \ref{fig:local_expl_eq_midcap} and \ref{fig:local_expl_bond_usd_short}, we visualize local Owen values for two sample fund descriptions. 

Then, for each class, we summed the Owen values of each token in the vocabulary over the entire test set, which is a measure of the global importance of each token for this class. On table \ref{table:global_explan}, we present the words with the 5 highest Owen value sums over the test set for a few sample frequent classes. While BERT does not classify based solely on the presence or absence of a given word, this measure of global importance allows us to explain which particular words might significantly increase the predicted probability of a given class as well as observe any other interesting patterns which explain the decision mechanism of the model. For example, for class \textit{Equity US}, \textbf{1000} and \textbf{500} are tokens with high importance. Most US Equity funds are benchmarked against S\&P 500 or Russell 1000 index. Thus, given that the benchmark is mentioned in the fund description, the model learned how to take it into account when classifying a fund as \textit{Equity US}. Overall, the model seems to attribute higher importance to words which are descriptive of the given class and do not occur often in examples of other classes (e.g. \textbf{emerging} for \textit{Equity Emerging Mkts Global}, \textbf{technology} for \textit{Equity Sector Information Tech}, \textbf{municipal} for \textit{Bond USD Municipal}). 

\textbf{Where does the model fail?} In Figure \ref{fig:conf_matrix} we present the confusion matrix for the top 30 most frequent classes. Given the class imbalance, we choose to show the proportion of correctly and incorrectly classified instances from the test set per class. The model predicts \textit{Equity US} funds with 87\% recall rate. However, as it can be observed from the first column of the confusion matrix, the model misclassified a significant proportion of other Equity-type funds (e.g. \textit{Equity US Sm \& Mid Cap, Equity Global, Equity US Income}). This is expected because \textit{Equity US} is a broad label which is by far the most frequent class in the training set (15\% of training instances). Given that other equity funds might have very similar descriptions to the general \textit{Equity US} but they are significantly underrepresented in the training set, the model overpredicts the majority class in some cases. This could potentially be mitigated by training the model after applying some class balancing techniques on the training set like undersampling/oversampling. 

Additionally, a significant proportion of misclassifications stems from \textit{Mixed Asset USD Aggressive} being mistaken for \textit{Mixed Asset USD Balanced} funds or the opposite. Usually, funds from these two categories share almost identical descriptions in terms of the language used but they differ in terms of the percentages allocated to different asset classes (aggressive funds allocate about 60-100\% of assets to equity funds while balanced funds allocate about 40-60\%). This is a common failure mode - in the absence of other words indicative of whether the fund is balanced or not, the model fails to do the math and infer the actual type of mixed fund using the percentages allocated across asset classes. The same failure mode is observed with \textit{Bond USD Short Term} funds being confused with \textit{Bond USD Medium Term} funds or the opposite. Again, most such funds have very similar descriptions with the difference only stemming from the target duration or maturity of the fund's constituents which is usually expressed with numbers (typically short-term funds have duration 1-3 years while medium term have 4-8 years, however there does not seem to be a hard-coded boundary between the two types). An interesting example to understand the decision mechanism of the model through the lens of Owen values is presented in Figure \ref{fig:local_expl_bond_usd_short}. The model mostly attends to the words \textbf{Month} and \textbf{Treasury Bill} (i.e. the more red, the higher the Owen value of the token) and predicts \textit{Bond USD Short Term} instead of the true label which was \textit{Bond USD Medium Term} in this case. The model's decision can be attributed to the fact that many short-term bonds funds invest in Treasury Bills which have duration shorter than one year, expressed typically in months. 

\vspace{-2mm}
\section{Conclusion}
In many problems in finance, labeled data is only a luxury. However, mutual fund categorization system provides a fascinating dataset where in addition to the input features arising from the real-world, each data-point is manually labeled by an expert based on their domain expertise and experience. The categorization systems have been continuously updated and fairly mature over the years. Such a dataset provides a unique opportunity for the ML algorithms to learn the embeddings from complex textual datasets from the experts' labels.

In the present work, we framed mutual fund categorization as a supervised multi-class classification problem, where the input textual data is a fund's investment strategy as depicted by the fund managers for an SEC form, and the target variable being Lipper Global categories. We explored several text classification methodologies including finetuning pre-trained BERT models on the aforementioned classification task, as well as text vectorization techniques such TF-IDF, Doc2Vec and Word2Vec combined with a logistic regression classifier.

We show that a TF-IDF baseline model with a logistic regression classifier is able to adequately reproduce the categories with top-2 accuracy ~80\% and F1 score ~66\%. Finetuned BERT was the best model, with FinBERT being a close second, with respect to all performance metrics. This is in accordance with our expectations demonstrating the significant advantage of transfer learning on NLP tasks.

In fact, top-k accuracy is quite high for fine-tuned BERT and FinBERT already for $k=3$ and $4$, meaning that the models can reproduce the categorizations with more than $92\%$ accuracy if the top-k classes with the highest predicted are considered. In other words, even in the case of misclassifications finetuned BERT assigned considerable probability to the ground truth label.

To further enhance these models one can, bring in additional text available in a fund’s prospectus. There are dedicated sections which throw more light on risks involved with investment products. Often this text tends to be more generic to satisfy regulation but could insted of useful to provide more contextual information to the language model. Previous work has explored predicting fund categories using structured data \cite{mehta2020machine}. One interesting direction of future work would be to combine the NLP based method presented here with the approach using structured data to improve classification accuracy.

\section*{Acknowledgement}
The views expressed here are those of the authors alone and not of BlackRock, Inc.

\bibliography{main}{}
\bibliographystyle{unsrt}

\end{document}